\documentclass[11pt,twoside,english]{article}
\usepackage{ae,aecompl}

\usepackage[T1]{fontenc}
\usepackage[latin9]{inputenc}
\usepackage[a4paper]{geometry}
\usepackage{array}
\usepackage{verbatim}
\usepackage{longtable}
\usepackage{booktabs}
\usepackage{url}
\usepackage{amsmath}
\usepackage{graphicx}
\usepackage{setspace}
\makeatletter

\providecommand{\tabularnewline}{\\}

\usepackage{fancyhdr} 
\usepackage{calc} 
\usepackage{color}
\usepackage{url}
\usepackage{sectsty}
\usepackage{fancybox}
\usepackage{psboxit}
\usepackage[font=small,format=plain,labelfont=bf,up,textfont=sl,up]{caption}
\usepackage{type1cm}
\allsectionsfont{\usefont{OT1}{phv}{bc}{n}\selectfont}

\definecolor{orangeSQ}{rgb}{1,0.57,0.4}
\definecolor{salmonSQ}{rgb}{1.0,0.91,0.82}
\definecolor{brownSQ}{rgb}{0.36,0.17,0.18}
\newcommand{\HRule}{\rule{\linewidth}{2pt}}
\def\blfootnote{\xdef\@thefnmark{}\@footnotetext}

\usepackage{float}
\usepackage{esint}

\@ifundefined{showcaptionsetup}{}{%
 \PassOptionsToPackage{caption=false}{subfig}}
\usepackage{subfig}
\makeatother

\usepackage{babel}

\begin{document}
\fancyheadoffset[LE,RO]{\marginparsep+30pt}

\renewcommand{\headrule}{{\color{orangeSQ}%
\hrule width\headwidth height\headrulewidth \vskip-\headrulewidth}}
\renewcommand{\headrulewidth}{2pt}

\renewcommand{\sectionmark}[1]{\markboth{#1}{}}  
\fancyhf{}
\fancyhead[LE,RO]{\bfseries \thepage}
\fancyhead[LO]{\bfseries \leftmark}  
\fancyhead[RE]{\bfseries Traders' collective portfolio optimization with transaction costs}
\fancyfootoffset[LE,RO]{\marginparsep+50pt}
\fancyfoot[RO]{
\includegraphics[height=0.7cm]{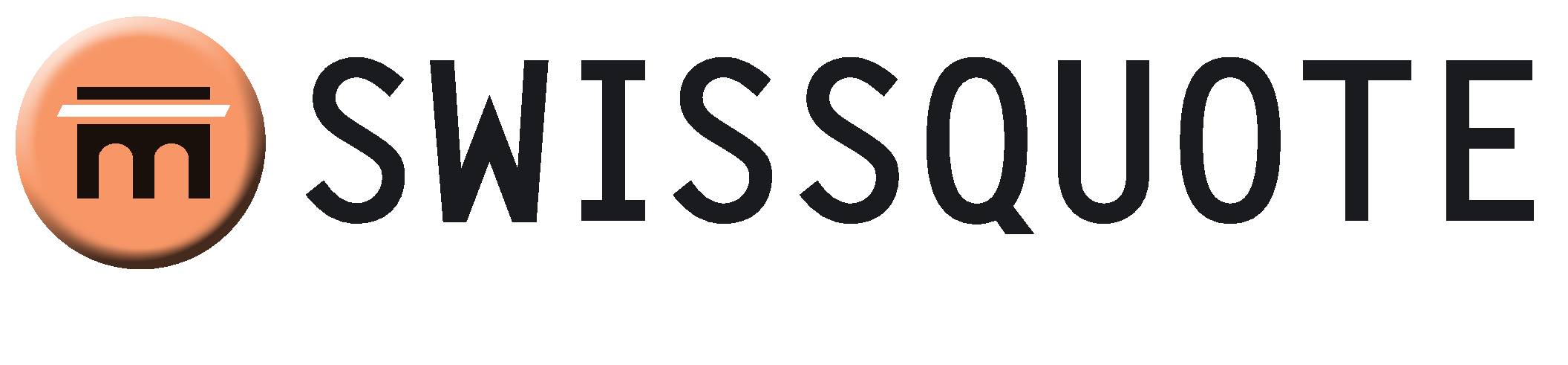} }

\fancypagestyle{plain}{%
\fancyhead{} 
\renewcommand{\headrulewidth}{0pt} 
}

\begin{titlepage}
\begin{center}     


\includegraphics[width=0.45\textwidth]{./logos/logo_sq_no_baseline_under_logo}\\[1cm] 
{\color{orangeSQ}\HRule} \\[1cm]

\textsf{\LARGE Turnover, account value and diversification of real
traders: \\evidence of collective portfolio optimizing behavior}\textsf{\huge \includegraphics[width=0.5cm]{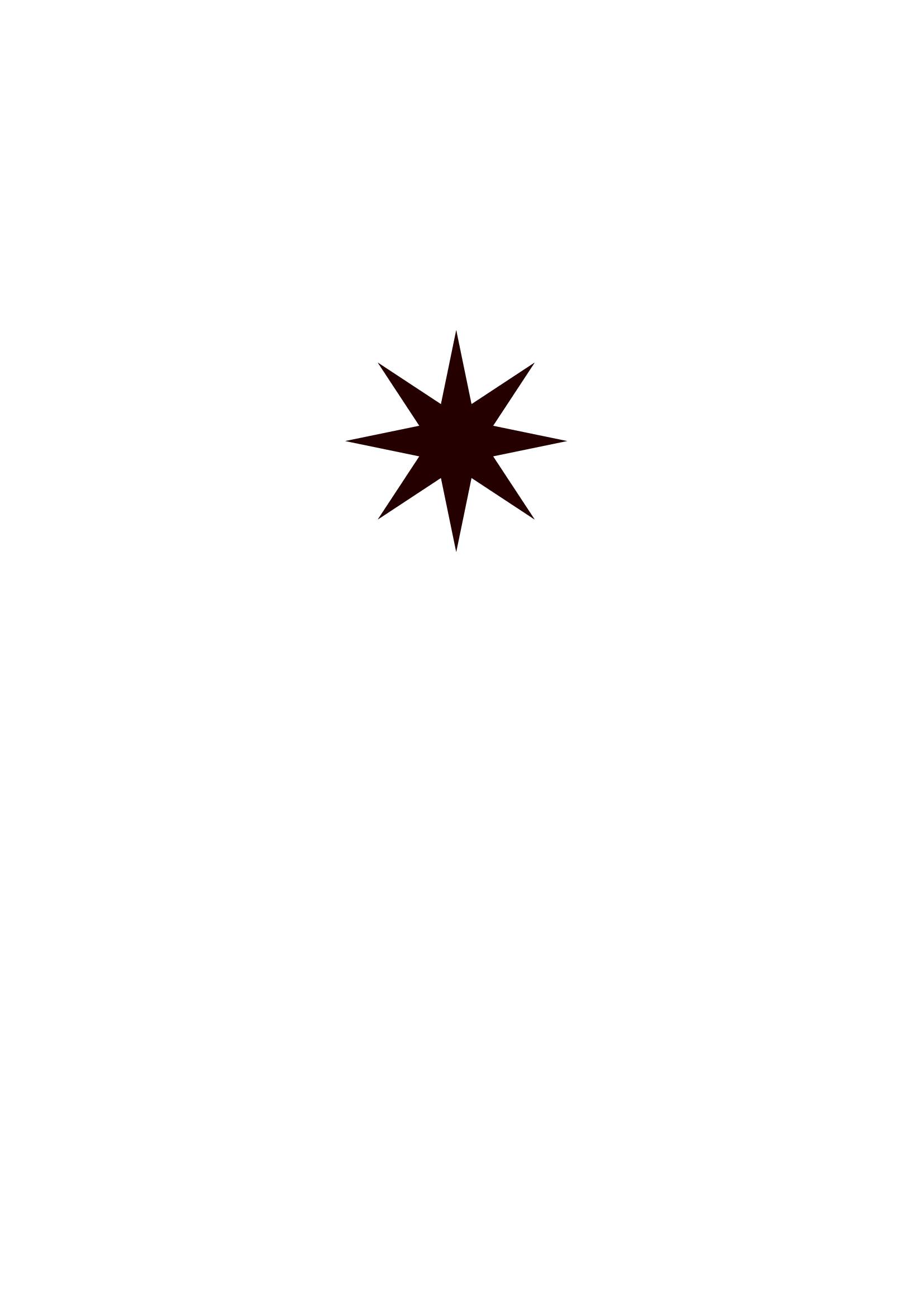}\blfootnote{\includegraphics[width=0.5cm]{./logos/brown_star}Early results in connexion with this project have been presented at the Fribourg Symposium (Oct. 2008, \url{unifr.ch/econophysics/symposium}), the Tokyo APFA7 Workshop (Feb. 2009, \url{thic-apfa7.com}), the EPFL Alliance Carrefour (Mar. 2009, \url{alliance-tt.ch/Carrefours}), and the Zurich Workshop on Complex Socio-Economic Systems (Jun. 2009, \url{soms.ethz.ch/workshop2009}).}}{\huge \par}

\vskip 0.7cm
{\color{orangeSQ}\HRule} \\[1.2cm]

{\LARGE David Morton de Lachapelle$^{1,2}$, Damien Challet$^{3}$}{\LARGE \par}

\vskip 1cm
\begin{minipage}{0.76\textwidth}

$^{1}$Swissquote Bank SA, \url{David.MortonDeLaChapelle@swissquote.ch}

$^{2}$LBS, Institute of Theoretical Physics, EPFL, \url{David.Morton@epfl.ch}

$^{3}$Physics Department, Fribourg University, \url{Damien.Challet@unifr.ch}

\end{minipage}

\vskip 2cm
\begin{minipage}{0.9\textwidth}
\small{

\textcolor{orangeSQ}{\textbf{\textit{\small Abstract~}}}~~~
Despite the availability of very detailed data on financial market,
agent-based modeling is hindered by the lack of information about
real trader behavior. This makes it impossible to validate agent-based
models, which are thus reverse-engineering attempts. This work is
a contribution to the building of a set of stylized facts about the
traders themselves. Using the client database of Swissquote Bank SA,
the largest on-line Swiss broker, we find empirical relationships
between turnover, account values and the number of assets in which
a trader is invested. A theory based on simple mean-variance portfolio
optimization that crucially includes variable transaction costs is
able to reproduce faithfully the observed behaviors. We finally argue
that our results bring into light the collective ability of a population
to construct a mean-variance portfolio that takes into account the
structure of transaction costs.}
\end{minipage}
\end{center}
\end{titlepage}

\setcounter{page}{2}

\section{Introduction}

\label{SEC:intro}

The availability of large data sets on financial markets is one of
the main reasons behind the number and variety of works devoted to
their analysis in various fields, and especially so in Econophysics
since physicists much prefer to deal with very large data sets. At
the macroscopic level, the analysis of millions of tick-by-tick data
points uncovered striking regularities of price, volume, volatility,
and order book dynamics (see~\cite{MantegnaStanley,BouchaudPotters,Daco,BouchaudFarmerLillo}
for reviews). Since these phenomena are caused by the behavior of
individual traders, news, and the interplay between the two, finding
a microscopic mechanism that allows agent-based models to reproduce
some of these stylized facts is an important endeavor meant to give
us insight on the causes for large fluctuations, be it herding~\cite{ContBouchaud},
competition for predictability~\cite{MGbook}, portfolio optimization
leading to market instability~\cite{MarsiliInstabMarkets}, or chaotic
transitions~\cite{BH97}.

Market phenomenology appears as a typical example of collective phenomena
to the eyes of statistical physicists. Thus, the temptation to regard
the numerous power-laws found in empirical works as signatures of
criticality is intense. But if the former are really due to a phase
transition, one wishes at least to know what the phases are, which
is hard to guess from the data alone. According to early herding theoretical
models~\cite{ContBouchaud}, the phase transition may lie in the
density of social communication and imitation, and is of percolation
type, thereby linking power-law distributed price and volume, criticality
and agent-behavior. The standard Minority Game~\cite{CZ97} has also
a single phase transition point where market predictability is entirely
removed by the agents, without any specular effect on price and volume;
on the other hand, grand-canonical MGs~ \cite{SZ99,J00,CCMZ00,CM03}
that allow the agents not to play have a semi-line of critical points
that do produce stylized facts of price, volume and volatility dynamics;
in the framework of statistical physics, the phase transition is due
to symmetry breaking, i.e., it is a transition between predictable
and perfectly efficient markets; this also suggests that the emergence
of large fluctuations is due to market efficiency.

There are of course many other possible origins of power-laws in financial
markets that have nothing to do with a second order phase transition.
The simplest mechanism is to consider multiplicative random walks
with a reflecting boundary~\cite{MZMPortfolio}. Long-range memory
of volatility is well-reproduced in agent-based models whose agents
act or do nothing depending on a criterion based on a random walk~\cite{BouchaudGiardina}.
Assuming pre-existing power-law distributed wealth, an effective theory
of market phenomenology links the distributions of price returns,
volume, and trader wealth~\cite{Gabaix_etalNature2003}. On the other
hand, markets are able to produce power-law distributed price returns
by simple mechanisms of limit order placement and removal without
the need for wealth inequality~\cite{CS02,FarmerLillo}. However,
in turn, one needs to explain why limit orders are placed in such
manner; the heterogeneity of time scales may provide an explanation
of order placement far away from best prices if power-law distributed
~\cite{LilloUtility}, but additional work is needed in order to
explain order placement near best prices, which causes these large
price moves. Finally, a recent simple model of investment with leverage
is able to reproduce some stylized facts~\cite{thurner-leverage}.

But mechanisms alone may not be sufficient to replicate the full complexity
of financial markets, as some part of it may lie instead in the heterogeneity
of the agents themselves. While the need for heterogeneous agents
in this context is intuitive (see e.g.~\cite{Arthur}), there is
no easily available data against which to test or to validate microscopically
an agent-based model. Even if it is relatively easy to design agent-based
models that reproduce some of the stylized facts of financial markets
(see e.g~\cite{LuxMarchesi,Mercatino,BH97,MGbook,Alfarano}), one
never knows if this is achieved for good reasons, except for volatility
clustering~\cite{BouchaudGiardina}: it is to be expected that real
traders behave sometimes at odds with one's intuition. Thus, without
data about the traders themselves, one is left with the often frustrating
and time-consuming task of reverse-engineering the market in order
to determine the good ingredients indirectly. Some progresses have
been made recently with the analysis of transactions in Spanish stock
market aggregated by brokers~\cite{vaglica2008scaling}, hence with
mesoscale resolution.

Data on trader behavior is found in the files of brokers, usually
shrouded in secrecy. But this lack of data accessibility is not entirely
to blame for the current ignorance of real-trader dynamics: researchers,
even when given access to broker data, have focused on trading gains
and behavioral biases, often with factor-based analyses (see e.g.
\cite{BarberTaiwan,BrokerNL,ghondi2010success}). 

We aim at providing a coherent picture of how various types of traders
behave and interact, making it possible for agent-based models to
rest on a much more solid basis. This paper is the first of a series
that will establish stylized facts about trader characteristics and
behavior. One of the most important aspects of these papers will be
to characterize the heterogeneity of the traders in all respects (account
value, turnover, trading frequency, behavioral biases, etc.) and the
relationships between these quantities in probability distribution,
not with factors. This paper is first devoted to the description of
the large data set that we use; it then focuses on the relationship
between trader account value, turnover per transaction and transaction
costs, both empirically and theoretically. We will show that while
the traders have a spontaneous tendency to build equally-weighted
portfolios, the number of stocks in a portfolio increases non-linearly
with their account value, which we link to portfolio optimisation
and broker transaction fee structure.

\section{Description of the data}

Our data are extracted from the database of the largest Swiss on-line
broker, Swissquote~Bank~SA (further referred to as Swissquote).
The sample contains comprehensive details about all the 19 million
electronic orders sent by 120'000 professional and non-professional
on-line traders from January 2003 to March 2009. Of these orders,
65\% have been canceled or have expired and 30\% have been filled;
the remaining 5\% percent were still valid as of the 31st of March
2009. Since this study focuses on turnover as a function of account
value, we chose to exclude orders for products that allow traders
to invest more than their account value, also called leveraging, i.e.,
orders to margin-calls markets such as the foreign exchange market
(FOREX) and the derivative exchange EUREX. The resulting sample contains
50\% of orders for derivatives, 40\% for stocks, and 4\% for bonds
and funds. Finally, 70\% of these orders were sent to the Swiss market,
20\% to the German market and about 10\% to the US market.

Swissquote clients consist of three main groups:\emph{ individuals},
\emph{companies}, and \emph{asset managers}. Individual traders, also
referred to as \emph{retail clients}, are mainly non-professional
traders acting for their own account. The accounts of companies are
usually managed by individuals trading on behalf of a company and,
as we shall see, behave very much like retail clients, albeit with
a larger typical account value. Finally, asset managers manage accounts
of individuals and/or companies, some of them dealing with more than
a thousand clients; their behavior differ markedly from that of the
other two categories of clients. %
\section{Results\label{sec:Results}}

\subsection{Account values}

\label{sub:wealth}

\begin{figure}
\noindent \begin{centering}
\includegraphics[scale=0.7]{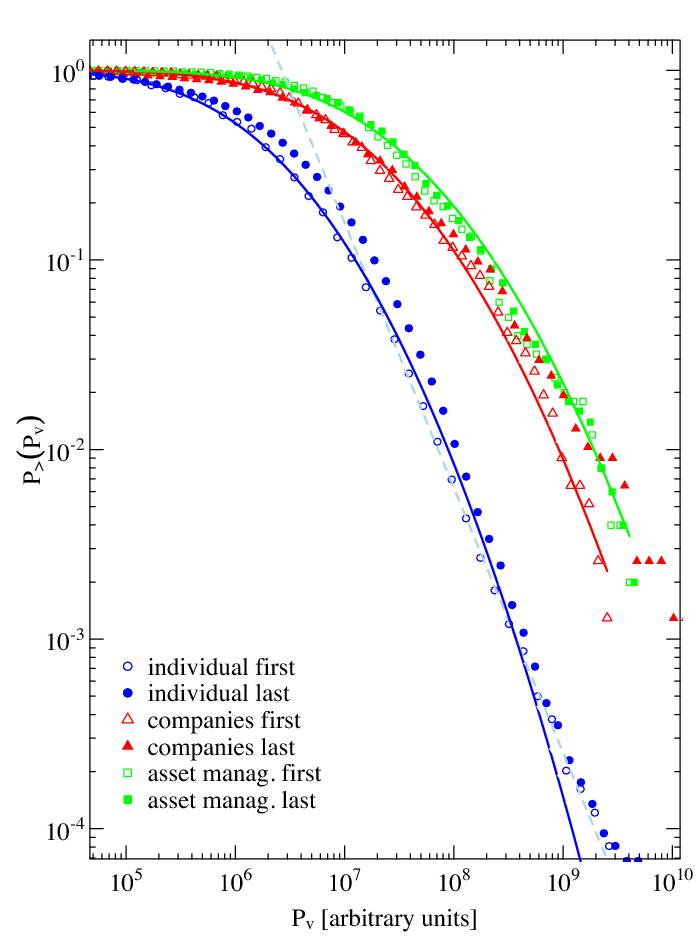}
\par\end{centering}

\caption{Reciprocal cumulative distribution function of the portfolio value
$P_{v}$ for the three categories of clients at the time of their
first (empty symbols) and last (filled symbols) transactions. Several
models have been fitted to the data by Maximum Likelihood Estimation
(MLE): the Student distribution (Pareto with plateau), the Weibull
(stretched exponential), and the log-normal distribution. The best
candidate, determined graphically and via bootstrapping the Kolmogorov
Smirnov test \cite{clauset2009powerlaw} was found to be the log-normal
distribution, which is the only one shown here for the sake of clarity.
The dashed line in light blue results from a MLE fit to the tail of
the individual traders with the Pareto distribution $p(x)\sim(x/x_{\min})^{-\gamma}$
(see section~\ref{sub:wealth}). \label{fig:cdf_W}}

\end{figure}

Numerous studies have been devoted to the analysis and modeling of
wealth dynamics and distribution among a population~(see \cite{yakovenko2007econophysics}
and references therein). The general picture is that in a population,
a very large majority lies in the exponential part of the reciprocal
cumulative distribution function, while the wealth of the richest
people is Pareto-distributed, i.e., according to a power-law. 

The account value of Swissquote traders is by definition the sum of
all their assets (cash, stock, bonds, derivatives, funds, deposits),
and denoted by $P_{v}$. In order to simplify our analysis, we compute
$P_{v}$ once per day after US markets close and take this value as
a proxy for the next day's account value. Figure~\ref{fig:cdf_W}
displays this distribution computed at the time of the first and last
transactions of the clients. Results are shown for the three main
categories of clients. Maximum likelihood fits to the tail of the
individual traders to the Pareto model $p(x)\sim(x/x_{\min})^{-\gamma}$
were performed using the BC$_{a}$ bootstrap method of~\cite{efron93bootstrap}
and determining the parameter $x_{\min}$ by minimizing the Kolmogorov-Smirnov
statistics as in \cite{clauset2009powerlaw}. Results are reported
in table \ref{tab:Pv_pareto}. 

\begin{table}
\centering{}\caption{Results of the fits of Pareto law $(x/x_{min})^{\gamma}$ to the account
value $P_{v}$ of individuals.\label{tab:Pv_pareto}}
\begin{tabular}{ccc}
\hline 
individuals & $\gamma$ & $x_{min}$\tabularnewline
\hline 
first transaction & $2.33\in[2.29,2.37]_{95}$ & $2.30\cdot10^{6}\in[1.99\cdot10^{6},2.59\cdot10^{6}]$\tabularnewline
\hline 
last transaction & $2.39\in[2.33,2.44]_{95}$ & $3.73\cdot10^{6}\in[3.15\cdot10^{6},4.29\cdot10^{6}]$\tabularnewline
\end{tabular}
\end{table}

The values of $\gamma$ are in line with the wealth distribution of
all major capitalistic countries (see~\cite{solomon2001power} for
a possible origin of Pareto exponents between 2.3 and 2.5). Thus the
retail clients are most probably representative of the Swiss population.
The account value distributions of companies and asset managers have
no clear power-law tails, in agreement with the results of a recent
model that suggests a log-normal distribution of mutual fund asset
sizes~\cite{farmerMutualFunds}. Consequently, figure~\ref{fig:cdf_W}
also reports a fit of the data to log-normal distributions $\ln N(\mu,\sigma^{2})$,
which approximate more faithfully $P_{>}(P_{v})$ than the Student
and the Weibull distributions for the three categories of clients,
except its extreme tail in the case of retail clients. 

\begin{table}
\caption{Parameter values and 95\% confidence intervals for the MLE fit of
the account values to the log-normal distribution $\ln N(\mu,\sigma^{2})$.
For each category of investors, the first and second row correspond
to the account value at the time of the first, respectively the last
transaction (see text). Note that portfolio values have been multiplied
by an arbitrary number for confidentiality reasons. This only affects
the value of $\mu$.\label{tab:Parameters-linear-model-T-Pv-1}}

\hfill{}\begin{tabular}{lcc}
\toprule 
 & $\mu$ & $\sigma$\tabularnewline
\midrule
individuals & $13.94\pm0.02$ & $2.87\pm0.01$\tabularnewline
 & $14.25\pm0.02$ & $2.01\pm0.01$\tabularnewline
\midrule
companies & $16.0\pm0.2$ & $2.0\pm0.1$\tabularnewline
 & $15.9\pm0.2$ & $2.4\pm0.1$\tabularnewline
\midrule
asset managers & $16.7\pm0.2$ & $1.8\pm0.1$\tabularnewline
 & $16.7\pm0.2$ & $2.0\pm0.1$\tabularnewline
\bottomrule
\end{tabular}\hfill{}
\end{table}

\subsection{Mean turnover}

\begin{figure}
\noindent \begin{centering}
\hfill{}\subfloat[Stocks]{\includegraphics[scale=0.65]{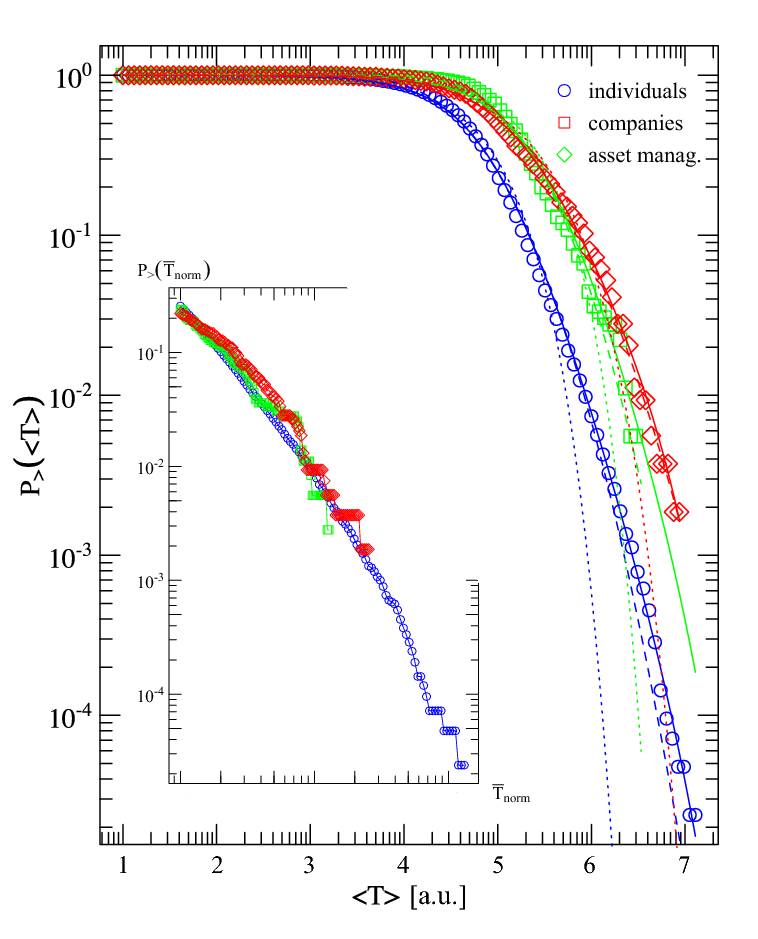}}\hfill{}\subfloat[Derivatives]{\includegraphics[scale=0.65]{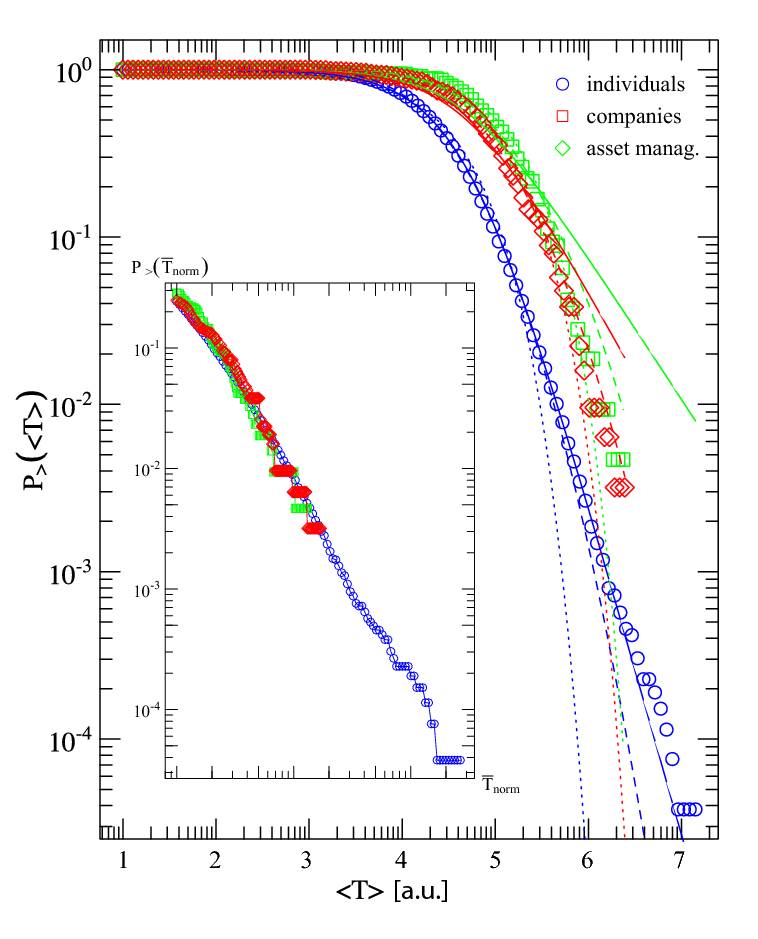}}\hfill{}
\par\end{centering}

\caption{Reverse cumulative distribution function of the mean turnover per
transaction for the three categories of clients, and for both stock
and derivative transactions. In the insets, the tail part of the RCDF
of $\left\langle T_{norm}\right\rangle =\left\langle T\right\rangle /\mbox{mean}(\left\langle T\right\rangle )$.
The solid curves are maximum likelihood fits to (\ref{eq:cdf_turnover_cutoff})
for stocks and (\ref{eq:cdf_turnover}) for derivatives. The dotted
lines are fits to the Weibull distribution and the dashed lines to
the log-normal distribution.\label{fig:RCDF_turnover}}

\end{figure}

The turnover of a single transaction $i$, denoted by $T_{i}$ is
defined as the price paid times the volume of the transaction and
does not include transaction fees. We have excluded the traders that
have leveraged positions on stocks, hence $T_{i}\le P_{v}$; more
generally one wishes to determine how the average turnover of a given
trader relates to his portfolio value. In passing, since $P(P_{v})$
has fat tails, the only way the distribution of $T$ can avoid having
fat tails is if the typical turnover is proportional to $\log(P_{v})$.
We denote by $\left<T\right>$ the mean turnover per transaction for
a given client over the history of his activities.

Figure~\ref{fig:RCDF_turnover} reports its reciprocal cumulative
distributions functions (RCDF) for stocks and derivatives for the
three categories of clients; all RCDFs have a first plateau and then
a fat tail. For stocks, the tails are not a pure power laws, but they
are for derivatives. Indeed, fitting the RCDFs with Weibull, log-normal
and Zipf-Mandelbrot distribution with an exponential cut-off, defined
as\begin{equation}
F_{>}^{(1)}(x)=\frac{c^{\gamma}e^{-\beta x}}{(c+x)^{\gamma}},\label{eq:cdf_turnover_cutoff}\end{equation}
clearly shows that the latter is the only one that does not systematically
underestimate the tail of the RCDF for stocks; estimated values of
$\beta$ and $\gamma$ given in table \ref{tab:fit_results_turn_cdf_stocks1}. 

The RCDFs related to the turnover of transactions on derivative products
have clearer power-law tails for retail clients, which we fitted with
a standard Zipf-Mandelbrot function, defined as \begin{equation}
F_{>}^{(2)}(x)=\frac{c^{\gamma}}{(c+x)^{\gamma}}.\label{eq:cdf_turnover}\end{equation}

The parameters estimated are to be found in table~\ref{tab:fit_results_turn_cdf_stocks1};
because of the power-law nature of this tail, fits with Weibull and
log-normal distributions are not very good in the tails. While the
decision process that allocates a budget to each type of product may
be essentially the same, the buying power is larger for derivative
products, which may explain the absence of a cut-off. Fits for companies
and asset managers is very difficult and mostly non-conclusive because
of unsufficient sample size; the good quality of the tail collapse
(see inset) tends to indicate that the three distributions are identical,
but we could not fit the RCDF of companies and asset managers with
(\ref{eq:cdf_turnover}); as reported in figure \ref{fig:RCDF_turnover}b,
log-normal distributions are adequate choices in these cases; since
the quality of the fits are poor, we do not report the resulting parameters\@.

\begin{table}
\caption{Results of the maximum likelihood fit of $P_{>}(\left<T\right>)$
with (\ref{eq:cdf_turnover_cutoff}) and (\ref{eq:cdf_turnover})
for the three categories of clients. The 95\% confidence intervals
reported in smaller character are computed by the biased-corrected
accelerated (BC$_{a}$) bootstrap method of~\cite{efron93bootstrap}.
\label{tab:fit_results_turn_cdf_stocks1}}

\noindent \centering{}\begin{longtable}{c|cc|c|}
\multicolumn{1}{c}{} & \multicolumn{2}{c}{Stocks (\ref{eq:cdf_turnover_cutoff})} & \multicolumn{1}{c}{Derivatives (\ref{eq:cdf_turnover})}\tabularnewline
\cline{2-4} 
 & $\gamma$ & $\beta\cdot10^{-6}$ & $\gamma$\tabularnewline
\hline
individuals & 1.97 & 0.98 & 1.98\tabularnewline
 & {\footnotesize {[}1.83,2.10{]}} & {\footnotesize {[}0.46,1.5{]}} & {\footnotesize {[}1.91,2.15{]}}\tabularnewline
\hline 
companies & 1.29 & 1.66 & -\tabularnewline
 & {\footnotesize {[}1.52,1.89{]}} & {\footnotesize {[}0.44,2.3{]}} & \tabularnewline
\hline 
asset managers & 1.93 & 0.91 & -\tabularnewline
 & {\footnotesize {[}1.47,2.93{]}} & {\footnotesize {[}-7.8,4.5{]}} & \tabularnewline
\hline
\end{longtable}
\end{table}

\newpage

\subsection{Mean turnover vs account value\label{sub:T-vs-Pv}}

\begin{figure}
\noindent \begin{centering}
\includegraphics[scale=0.7]{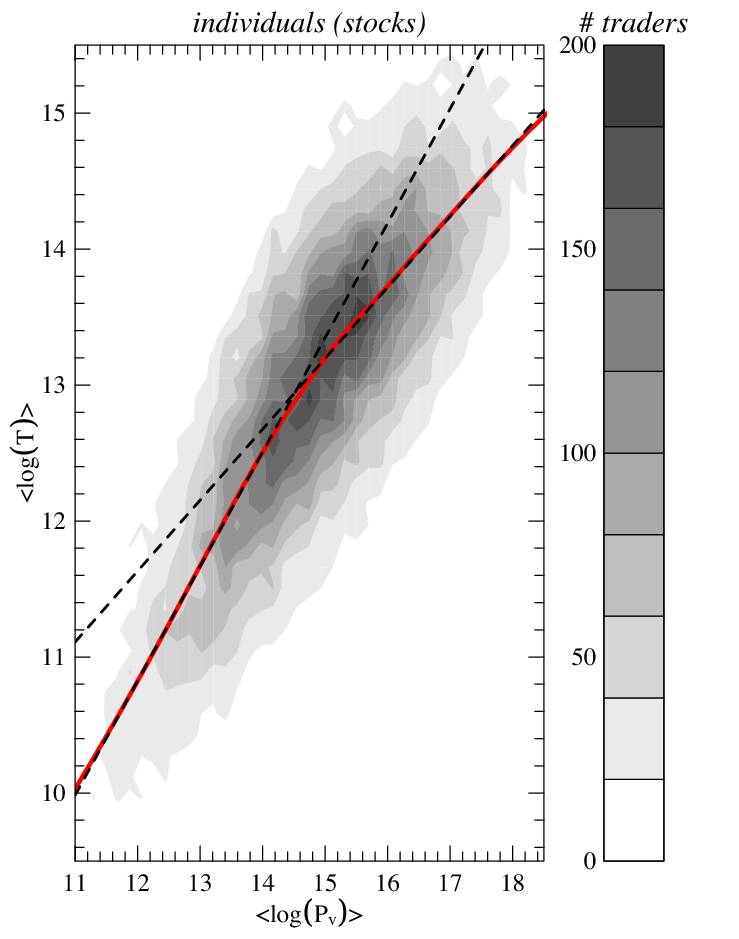}
\par\end{centering}

\caption{\label{fig:density_turnover_vs_wealth-1}Density plot of the average
$\log T$ vs the average $\log P_{v}$, robust non-parametric fit
(red line), and linear fits (dashed lines)}

\end{figure}

\begin{figure}
\noindent \begin{centering}
\hfill{}\subfloat[]{\includegraphics[scale=0.85]{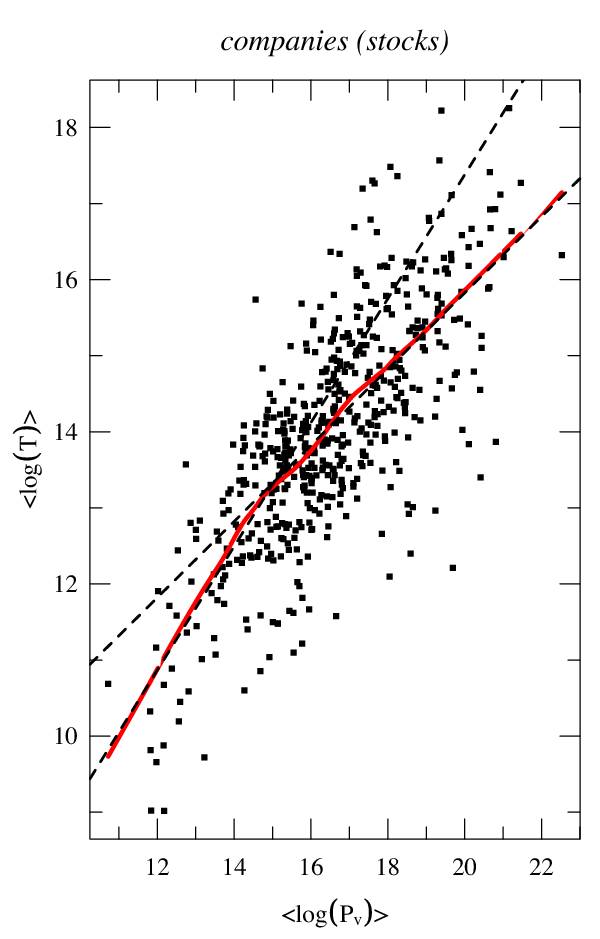}

}\hfill{}\subfloat[]{\includegraphics[scale=0.85]{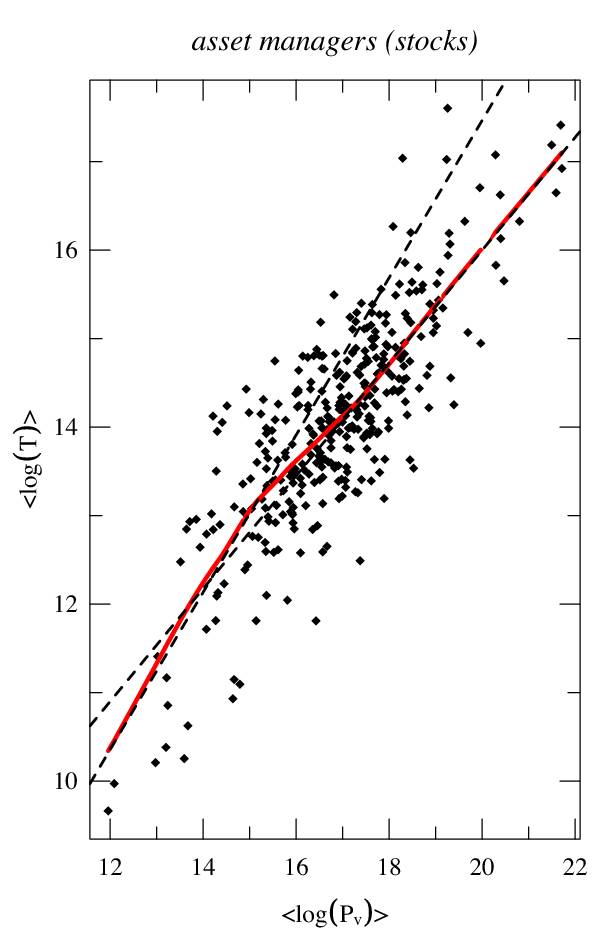}

}\hfill{}
\par\end{centering}

\caption{Density plot of the average $\log T$ vs the average $\log P_{v}$,
robust non-parametric fit (red line), and linear fits (dashed lines)\label{fig:T_vs_W_cloud-1}}

\end{figure}

The relationship between $\left<T\right>$ vs $\left<P_{v}\right>$
is important as it dictates what fraction of their investable wealth
the traders exchange in markets. We first produce a scatter plot of
$\left<\log T\right>$ vs $\left<\log P_{v}\right>$ (figure~\ref{fig:T_vs_W_cloud-1}).
In a log-log scale plot, it shows a cloud of points that is roughly
increasing. A density plot is however clearer for retail clients as
there are many more points (figure~\ref{fig:density_turnover_vs_wealth-1}).

{}These plots make it clear that there are simple relationships between
$\log T$ and $\log P_{v}$. A robust non-parametric regression method
\cite{cleveland1988} reveals a double linear relationship between
$\left\langle \log T\right\rangle $ and $\left\langle \log P_{v}\right\rangle $
for all three categories of investors (see figures~\ref{fig:T_vs_W_cloud-1}
and \ref{fig:density_turnover_vs_wealth-1}):\begin{equation}
\left\langle \log T\right\rangle =\beta_{x}\left\langle \log P_{v}\right\rangle +a_{x}\label{eq:logET_logEW_relation}\end{equation}
where $x=1$ when $\left\langle \log P_{v}\right\rangle <\Theta_{1}$
and $x=2$ when $\left\langle \log P_{v}\right\rangle >\Theta_{2}$.
Fitted values with confidence intervals are reported in table~\ref{tab:Parameters-linear-model-T-Pv}.

\begin{table}
\caption{Parameter values and 95\% confidence intervals for the double linear
model (\ref{eq:T_W_relation}). For each category of investors, the
first and second row correspond respectively to $\left\langle \log P_{v}\right\rangle \le\Theta_{1}$
an $\left\langle \log P_{v}\right\rangle \ge\Theta_{2}$. For confidentiality
reasons, we have multiplied $P_{v}$ and $T$ by a random number.
This only affects the true values of $a_{x}$ and $\Theta$ in the
table.\label{tab:Parameters-linear-model-T-Pv}}

\hfill{}\begin{tabular}{lccccc}
\toprule 
 & $\beta_{x}$ & $a_{x}$ & $\xi$ & $\Theta$ & $R^{2}$\tabularnewline
\midrule
individuals & $0.84\pm0.02$ & $0.73\pm1.25$ & 0.71 & 14 & 0.52\tabularnewline
 & $0.54\pm0.01$ & $5.07\pm0.15$ & 0.77 & 14.5 & 0.40\tabularnewline
\midrule
companies & $0.81\pm0.13$ & $1.12\pm8.17$ & 0.88 & 15.5 & 0.47\tabularnewline
 & $0.50\pm0.07$ & $5.82\pm1.65$ & 1.00 & 15.6 & 0.33\tabularnewline
\midrule
asset managers & $0.89\pm0.20$ & $-0.31\pm0.76$ & 0.62 & 15.5 & 0.52\tabularnewline
 & $0.63\pm0.08$ & $3.28\pm5.78$ & 0.62 & 16.5 & 0.46\tabularnewline
\bottomrule
\end{tabular}\hfill{}
\end{table}

This result is remarkable in two respects: (i) the double linear relation,
not obvious to the naked eye, separates investors into two groups
{}(ii) the ranges of values where the transition occurs is very similar
across the three categories of traders.

The relationships above only applies to averages over all the agents.
This means that there are some intrinsic quantities that make all
the agents deviate from this average line. Detailed examination of
the regression residuals show that the latter are for the most part
(i.e.~more than 95\%) normally distributed with constant standard
deviations $\xi_{x}$ and that the residuals deviating from the normal
distributions are not fat-tailed. This directly suggests the simple
relation for individual traders\begin{equation}
T^{i}=e^{a_{x}+\delta^{i}a_{x}}(P_{v}^{i})^{\beta_{x}}\le e^{\Theta_{x}}\label{eq:T_W_relation}\end{equation}

where $T^{i}$ and $P_{v}^{i}$ are respectively the turnover and
portfolio value of investor $i$, and $\delta^{i}a_{x}$ are i.i.d.
$N(0,\xi_{x}^{2})$ idiosyncratic variations independent from $P_{v}$
that mirror the heterogeneity of the agents. As we shall see, portfolio
optimization with heterogeneous parameters yields this precise relationship.

\subsection{Turnover rescaled by account value\label{sub:Turnover-rescaled}}

Let us now measure the typical fraction of wealth exchanged in a single
transaction, defined as $Q=\left<\frac{T}{P_{v}}\right>$. Since the
inverse of this ratio is an indirect (and imperfect) proxy of the
number $N$ of assets that a trader owns, it also indicates how well
diversified his investments are, hence, it can be viewed a simple
proxy of the risk profiles of the agents.

\subsubsection{data}

\label{sub:Turnover-rescaled-by}
\begin{figure}
\noindent \begin{centering}
\hfill{}\subfloat[]{\noindent \begin{centering}
\includegraphics[scale=0.65]{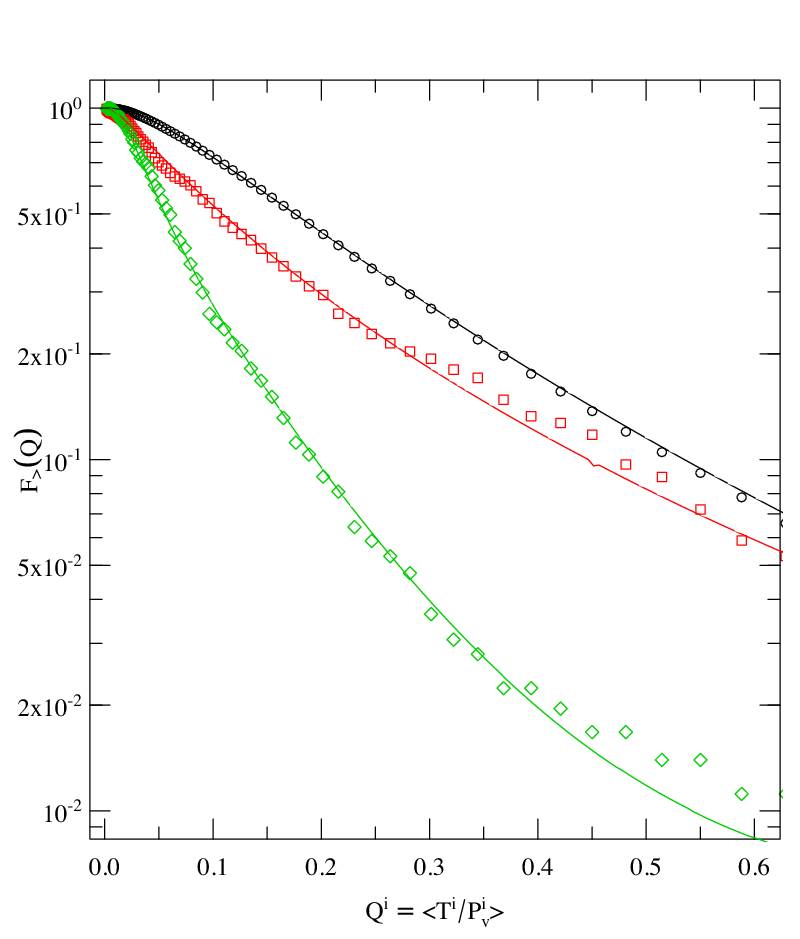}
\par\end{centering}

}\hfill{}\subfloat[]{\includegraphics[scale=0.65]{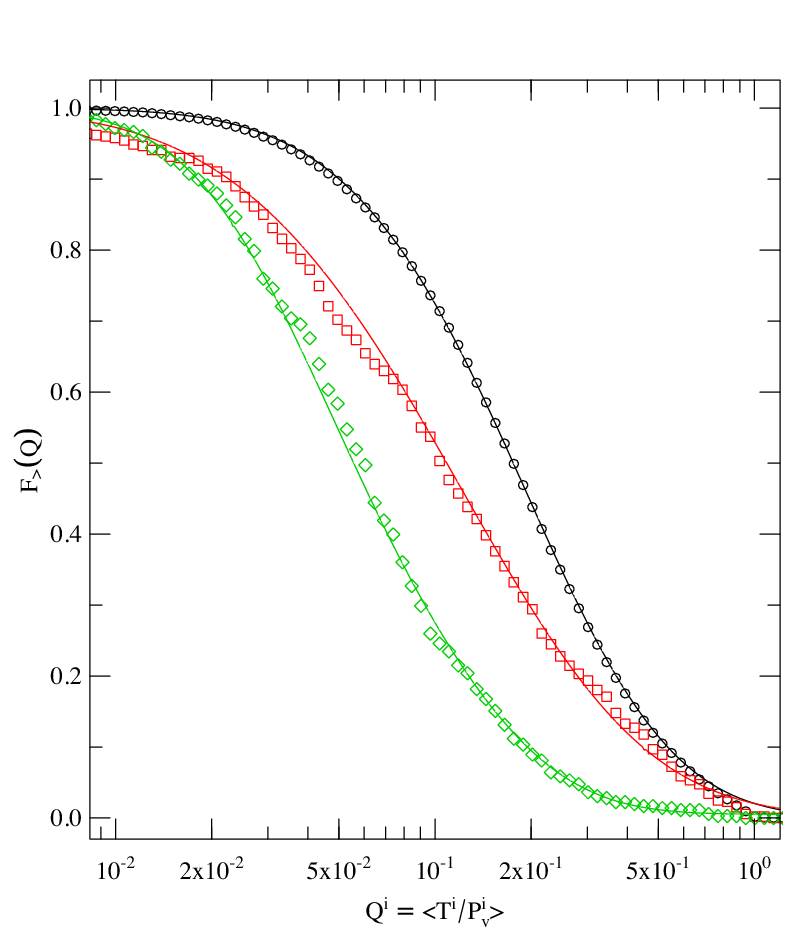}}\hfill{}
\par\end{centering}

\caption{Reverse cumulative distribution function of $Q=\left<\frac{T}{P_{v}}\right>$,
the mean ratio of the turnover over the portfolio value for individual
traders (black), companies (red) and asset managers (green). Left
plot is in lin-log scale and right plot is in log-lin scale. Solid
lines come from theoretical predictions of section~\ref{sub:Turnover-rescaled-by2}.
\label{fig:scaled_mean_turnover}}

\end{figure}

Figure~\ref{fig:scaled_mean_turnover}
shows that the distributions look exponential to a naked eye for about
90\% of the individuals and nearly 80\% of the companies, while that
of the asset managers is rapidly more complex that a simple exponential.
We derive exact relationships for this quantity in subsection~\ref{sub:Turnover-rescaled-by2}
that show that these distributions are in fact not exponential but
log-normal.

The resulting picture is that only a small fraction of customers trade
a large fraction of their wealth on average. Interestingly, these
figures show a clear difference between the three categories of clients.
As discussed above, figure~\ref{fig:scaled_mean_turnover} roughly
reflects the risk profile of the different types of customers: less
than 10\% of asset managers trade on average more than 20\% of their
clients' capital in a single transaction; this rises to 30\% for companies,
and 45\% for retail clients. Note however that despite the fact that
the account values of companies and asset managers are comparable,
companies tend to have a $Q$ closer to that of the individuals; this
suggests either that companies hold a smaller $N$ than asset managers
for the same account value, or that asset managers tend to make smaller
adjustments to the quantities of assets. 

\subsubsection{theory\label{sub:Turnover-rescaled-by2}}

Since we know the distributions of $T$, $P_{v}$ and their relationship,
we are in a position to derive analytical expressions for $Q^{i}=\left\langle \frac{T(t)}{P_{v}(t)}\right\rangle $
of investor $i$. The distribution of $Q$ across the population of
on-line investors can be easily found using (\ref{eq:T_W_relation})
and the distribution of $P_{v}$. Let $P_{T,P_{v}}(t,p_{v})$ denote
the joint distribution of $T$ and $P_{v}$:\begin{equation}
P_{Q=\frac{T}{P_{v}}}(q)=\int_{0}^{\infty}p_{v}P_{T,P_{v}}(qp_{v},p_{v})\, dp_{v}=\int_{0}^{\infty}p_{v}P_{T|P_{v}}(qp_{v}|p_{v})P_{P_{v}}(p_{v})\, dp_{v}.\label{eq:distribution_of_Q}\end{equation}

Let us now assume for the sake of clarity that $T=e^{a+\delta a}P_{v}^{\beta}$.
Given $P_{v}$, the turnover $T$ follows a log-normal distribution
with mean $\log p_{v}+a$ and variance $\xi^{2}$. Substituting $P_{T|P_{v}}(t|p_{v})=\ln N\left(\log p_{v}+a,\xi^{2}\right)$
in (\ref{eq:distribution_of_Q}) leads after some simplifications
to

\begin{equation}
P_{Q}(q)=\int_{0}^{\infty}\frac{1}{\sqrt{2\pi\xi^{2}}q}\exp\left(-\frac{\left(\log(qp_{v}^{1-\beta})-a\right)^{2}}{2\xi^{2}}\right)P_{P_{v}}(p_{v})\, dp_{v},\label{eq:distribution_of_Q_integral}\end{equation}

and

\begin{equation}
F_{Q}(q)=\int_{0}^{q}P_{Q}(x)\, dx=\int_{0}^{\infty}\frac{1}{2}\mbox{erfc}\left(\frac{a-\log(qp_{v}^{1-\beta})}{\sqrt{2}\xi}\right)P_{P_{v}}(p_{v})\, dp_{v},\label{eq:cdf_of_Q_integral}\end{equation}

where $\mbox{erfc}(x)=\frac{2}{\sqrt{\pi}}\int_{x}^{\infty}e^{-y^{2}}\, dy$
is the complementary error function. As expected, when $\beta=0$
(i.e.~$T$ and $P_{v}$ are independent), we recover the product
of the two marginal distributions. On the other hand, when $\beta=1$,
i.e., when $T$ is proportional to $P_{v}$, $P_{Q}(q)=\ln N\left(a,\xi^{2}\right)$,
which is the distribution of the factor $e^{a+\delta a}$. For other
values of $\beta$ the functions $P_{Q}$ and $F_{Q}$ cannot be determined
analytically unless $P_{P_{v}}$ takes a particular form as shown
below. However, the moments of $P_{Q}(q)$ can be arranged in a simpler
form:\begin{equation}
E(q^{n})=\int_{0}^{\infty}q^{n}P_{Q}(q)dq=e^{na+\frac{1}{2}n^{2}\xi^{2}}\int_{0}^{\infty}\frac{1}{p_{v}^{n(1-\beta)}}P_{P_{v}}(p_{v})\, dp_{v},\label{eq:moments_of_order_s}\end{equation}

that is, the (log-normal) moments of $T/P_{v}$ times an integral
term smaller or equal to $1$ (because in practice $P_{P_{v}}(p_{v})>1$)%
\footnote{Mathematically, all the moments of $Q$ always exist since $\beta\le1$
and $P_{v}(p_{v})$ must decay faster than $p_{v}^{-1}$ to be a valid
distribution.%
}. Hence, the relation $E(q^{n})\le e^{na+\frac{1}{2}n^{2}\xi^{2}}$
with equality when $\beta=1$ holds for any distribution of the account
value $P_{v}$.%
{}

In section~\ref{sub:wealth}, we have shown that the distribution
of $P_{v}$ is well-approximated by a log-normal distribution. This
particular choice of distribution makes the previous integrals analytically
tractable. Indeed, with $P_{P_{v}}=\ln N(\mu,\sigma^{2})$ straight
integration of (\ref{eq:distribution_of_Q_integral}) leads to $P_{Q}=\ln N(M,S^{2})$,
where $M=a-(1-\beta)\mu$ and $S^{2}=\xi^{2}+(1-\beta)^{2}\sigma^{2}$.
This simple result has some practical interest: given the distribution
parameters and the coupling factor $\beta$, one can draw realistic
$q$ factors for agent-based modeling as $Q=e^{M+SX}$, where $X$
is $\mathcal{N}(0,1)$ distributed. Furthermore, in the next section,
we show how the value of $\beta$ may be inferred from the transaction
cost structure, which decreases the number of parameters to four.

Figure~\ref{fig:scaled_mean_turnover} confirms the validity of the
above theoretical results, once expanded to the case of a bi-linear
relation between $T$ and $P_{v}$. It is noteworthy that the continuous
lines are no fits on empirical $q$ factors, but use instead the results
of the separate fits on the turnover and account distributions.

\section{The influence of transaction costs on trading behavior: optimal mean-variance
portfolios\label{sec:portfolo optimization model}}

\begin{figure}
\noindent \begin{centering}
\includegraphics[scale=0.7]{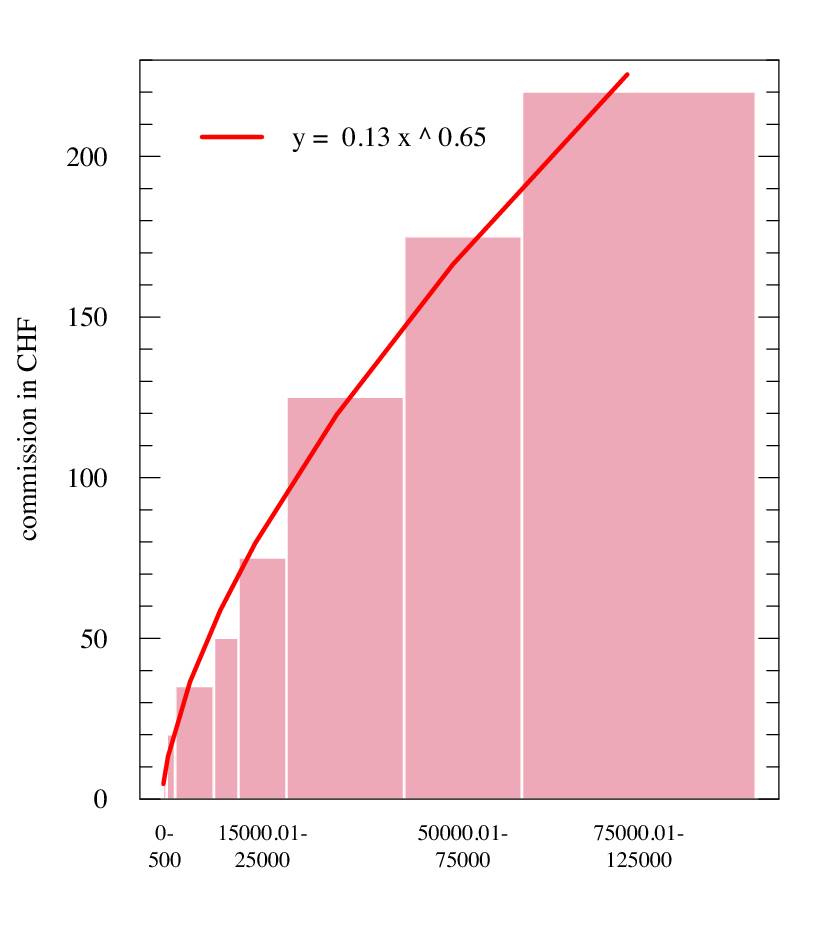}
\par\end{centering}

\caption{Swissquote fee curve for the Swiss stock market. Commissions based
on a sliding scale of costs are common practice in the world of on-line
finance. The red line results from a non-linear fit to equation~\ref{eq:cost_function}.
Parameter values are $C=0.13\in[0.05,0.5]_{95}$ and $\delta=0.63\in[0.5,0.74]_{95}$,
where the 95\% confidence intervals are obtained from the BC$_{a}$
bootstrap method of~\cite{efron93bootstrap}.\label{fig:Swissquote-fee-curve}}

\end{figure}

Apart from risk profiles, education, and typical wealth, the differences
in the turnover as a function of wealth observed above between the
three populations of traders may also lie in the difference of their
actual transaction cost structure. Swissquote current standard structure
for the Swiss market (its shape is very similar for European and US
markets) is shown in figure~\ref{fig:Swissquote-fee-curve}; it is
a piece-wise constant, non-linear looking function. Fitting all segments
to equation~\ref{eq:cost_function} gives $\delta=0.63\in[0.5,0.74]_{95}$.
The fee structure of most brokers is not set in stone and can be negotiated.
A frequent request is to have a flat fee, i.e. a fixed cost per transaction
corresponding to a constant function. Since quite clearly the negotiation
power of large clients or of clients that carry out many transactions
is more important, asset managers are more likely to obtain a more
favorable fee structure than basic retail clients. 

Since buying some shares of an asset is the result of unconscious
or calculated portfolio construction process, one first needs a theoretical
reference point with which to compare the population characteristics
as measured in the previous subsection. In other words, we shall use
results from portfolio optimization theory with non-linear transaction
cost functions to understand the results of the previous subsection.

Quite curiously, all analytical papers in the literature on optimal
portfolios either neglect transaction costs or assume constant or
linear transaction cost structures; non-linear structures are tackled
numerically; thus, we incorporate the specific non-linear transaction
cost structure faced by the traders under investigation in the classic
one-shot portfolio optimization problem studied by Brennan \cite{brennan1975optimal},
who restricted its discussion to fees proportional to the number of
securities, in other words, a flat fee per transaction.

Building optimal mean-variance stock portfolios consists for a given
agent in selecting which stock to invest in and in what proportion
by maximizing the expected portfolio growth, usually called return,
while trying to reduce the resulting a priori risk. One cost function
that corresponds to such requirements is

\begin{equation}
L_{\lambda}(R)=\lambda E(R)\mbox{-Var}(R),\label{eq:functional_L}\end{equation}

where $R$ is the stochastic return of the portfolio over the investment
horizon (e.g., one month, one year) and $\lambda$ tunes the trade-off
between risk and return; as such, it can be interpreted as a measure
of an investor's attitude towards risk: the larger $\lambda$ , the
more risk-adverse the investor. 

The return of the portfolio can be decomposed into contributions from
risky assets (stocks, derivatives, etc.), the interests of the amount
kept in cash, and the total relative cost of broker commission, which
we denote as $R=R^{risky}+R^{cash}-R^{cost}$. Mathematically,
\begin{itemize}
\item $R^{risky}=\Sigma_{i=1}^{N}x_{i}R_{i}$, where $R_{i}$ is the return
of stock $i$ over this horizon, $x_{i}$ is the fraction of the total
wealth invested in this stock, and $N$ is the total number of investable
assets; we shall denote the total fraction of wealth invested in risky
assets by $x=\sum_{i=1}^{N}x_{i}$;
\item $R^{cash}=(1-x)r$, where $r$ is the interest rate;
\item $R^{cost}=\frac{\sum_{i=1}^{N}F(x_{i}P_{v})}{P_{v}}(1+r)$, where
$F(x)$ is the amount charged by a broker to exchange an amount $x$
of cash into shares or vice-versa.
\end{itemize}
The focus of this section is to derive explicit relationships between
$F$, the number of assets to hold in a portfolio, and the account
value $P_{v}$. Whereas previous works only considered special cases
for $F$ that are not compatible with the fees structure of Swissquote,
we need to introduce a cost function that can accommodate all the
standard broker commission schemes. The two extreme cases are i) flat-fee
per transaction, i.e., a fixed cost that does not depend on the amount
exchanged ii) a proportional scheme, possibly with a maximum fee.
Swissquote's standard scheme stands in between and is well approximated
by a power-law with a maximum fee $F_{max}$. We hence choose \begin{equation}
F(x_{i}P_{v})=\min\left(C(x_{i}P_{v})^{\delta},F_{max}\right),\label{eq:cost_function}\end{equation}
where $\delta$ interpolates between a flat-fee ($\delta=0$), as
in \cite{brennan1975optimal}, and a proportional scheme ($\delta=1$)
via a power-law, and $C$ is a constant.

Following the well-known one-factor model of Sharpe \cite{sharpe1964capital},
we assume that the return of asset $i$ follows the global market's
return $R_{M}$ with an idiosyncratic proportionality factor $\beta_{i}$.
More specifically, \begin{equation}
R_{i}=\beta_{i}(R_{M}-r)+r+\varepsilon_{i},\label{eq:sharpe}\end{equation}
where $\varepsilon_{i}$ is an uncorrelated white noise $E(\varepsilon_{i})=E(\varepsilon_{i}\varepsilon_{j})=E(R_{M}\varepsilon_{i})=0.$
This equation means that the systematic idiosyncratic part of $R_{i}$
only applies to the return above the risk-free interest rate, also
called market risk premium.

This completely specifies the functional $L_{\lambda}$. Returning
to (\ref{eq:functional_L}), one first computes the expectation and
variance of the portfolio return:\begin{eqnarray}
E(R) & = & \sum_{i=1}^{N}x_{i}E(R_{i})+(1-x)r-\frac{\sum_{i=1}^{N}F(x_{i}P_{v})}{P_{v}}(1+r),\nonumber \\
 & = & (E(R_{M})-r)\sum_{i=1}^{N}x_{i}\beta_{i}+r-\frac{(1+r)C}{P_{v}^{1-\delta}}\sum_{i=1}^{N}x_{i}^{\delta},\label{eq:expected_portfolio_return}\end{eqnarray}
and\begin{eqnarray}
\mbox{Var}(R) & = & \mbox{Var}(R^{risky})\nonumber \\
 & = & \mbox{Var}(R_{M})\sum_{i=1}^{N}(x_{i}\beta_{i})^{2}+\sum_{i=1}^{N}x_{i}^{2}\mbox{Var}(\varepsilon_{i}).\label{eq:variance_portfolio_return}\end{eqnarray}
Note that, since here the risk-free rate is non-random, the portfolio
variance is independent of both the risk-free investment and broker
commission; this does not hold for the expected return.

In principle, the functional $L$ depends on  $N$, the number of
assets in the portfolio, $\lambda$ the risk parameter, and $x_{i}$
the fraction of account value to invest in risky product $i$. Assuming
that $x_{i}$ is constant for all $i$ (i.e.~equally-weighted allocation),
we are left with only three parameters since $x_{i}=x/N$. Thus, from
the optimization of the resulting functional one can obtain a relationship
between any two of these parameters. We are mostly interested in $N$
as a function of $x$.

\subsection{Non-linear relationship between account value and number of assets\label{sub:Optimal-equally-weighted-portfolio}}

\begin{figure}
\hfill{}\includegraphics[scale=0.55]{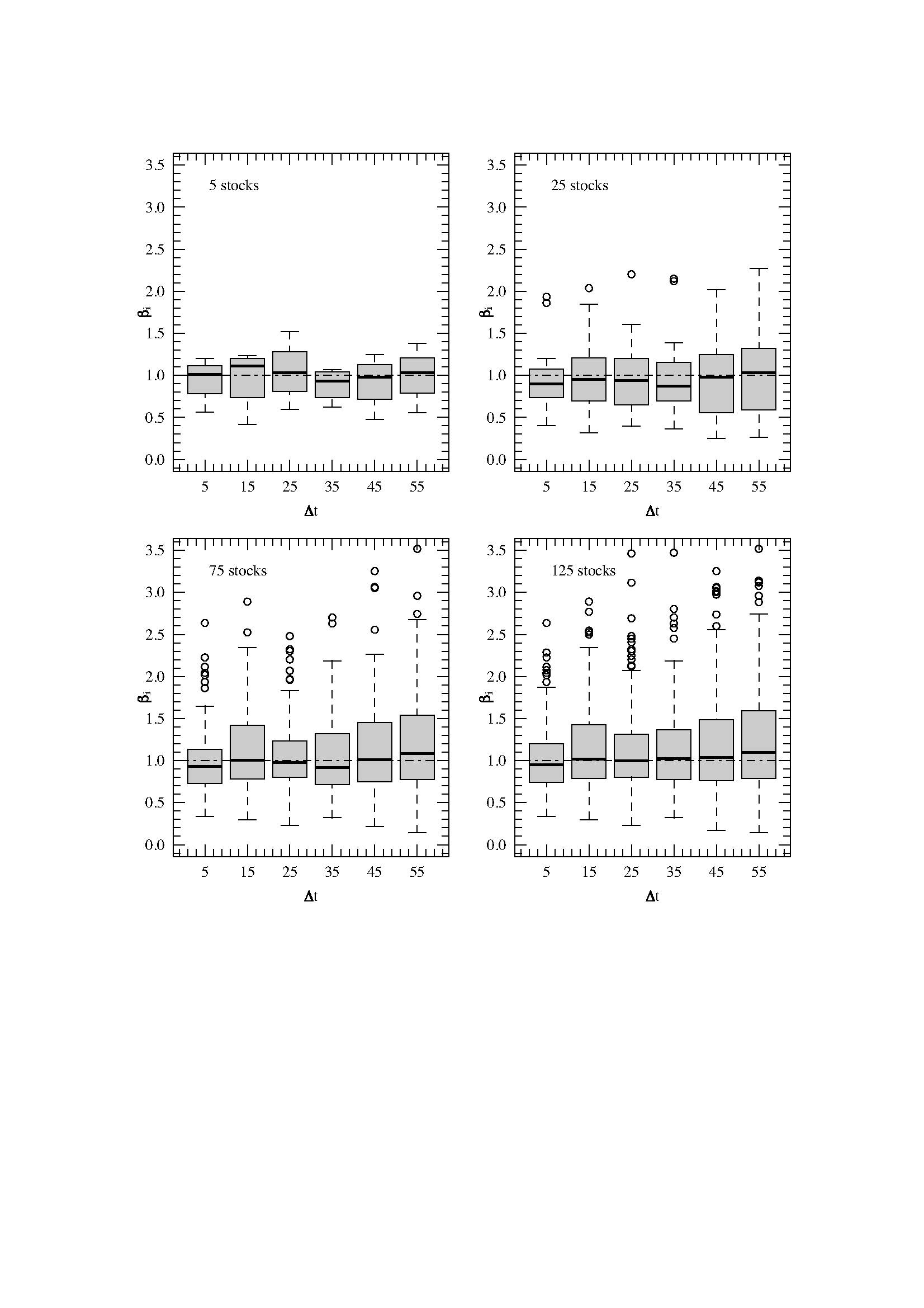}\hfill{}

\caption{Box-plot of empirical $\beta$s obtained from the regression of several
US stocks on the S\&P500. The observation period covers 2001 to 2008
and returns are computed on various time horizons $\Delta t$ (in
days). Results show that $\bar{\beta}=\frac{1}{N}\sum_{i=1}^{N}\beta_{i}\approx1$
for all values of $\Delta t$ and (even small) $N$, consistently
with the homogeneous assumption of section~\ref{sub:Optimal-equally-weighted-portfolio}.\label{fig:Boxplot_empirical_betas}}

\end{figure}

We will first assume that agents seek the optimal fraction of their
account value $x^{*}$ to invest in $N$ securities---\emph{$N$ being
known}---given the risk free rate $r$ and broker commission $F(x_{i}W)$.
The optimal solution is simply obtained by setting $x_{i}=x/N$ in
(\ref{eq:expected_portfolio_return}) and (\ref{eq:variance_portfolio_return}),
and by equating to zero the derivative of (\ref{eq:functional_L})
with respect to $x$. This leads to the following transcendental equation
for $x^{*}$: \begin{equation}
x^{*}=\frac{\lambda}{2}\frac{\bar{\beta}(E(R_{M})-r)-\delta(1+r)C(\frac{N}{x^{*}P_{v}})^{1-\delta}}{\bar{\beta}^{2}\mbox{Var}(R_{M})+\frac{1}{N}\overline{\mbox{Var}}(\varepsilon)},\label{eq:optimal_fraction_equal_weight}\end{equation}
where $\bar{\beta}=\frac{1}{N}\sum_{i=1}^{N}\beta_{i}$ and $\overline{\mbox{Var}}(\varepsilon)=\frac{1}{N}\sum_{i=1}^{N}\mbox{Var}(\varepsilon_{i})$
is the mean idiosyncratic volatility. Provided the investor risk tolerance
$\lambda$ has been reliably estimated, which is usually a complex
task~\cite{RePEc:eee:joepsy:v:29:y:2008:i:2:p:226-252}, and that
Sharpe model is adequate, (\ref{eq:optimal_fraction_equal_weight})
can be used directly in a real-world portfolio optimization problem.
The $\beta_{i}$ and $\varepsilon_{i}$ are then obtained by regressing
the returns of all the stocks with (\ref{eq:sharpe}); the optimal
solution is expected to be reliable in the absence of significant
residual correlations between $\varepsilon_{i}$ and $\varepsilon_{j}$.
In the more common situation where $\lambda$ is unknown, one can
derive a second equation for the optimal number of securities under
the assumption that portfolios are sufficiently homogeneous, or that
the investment horizon is long enough so as to have $\bar{\beta}$
and $\overline{\mbox{Var}}(\varepsilon)$ independent from $N$. As
shown in figure~\ref{fig:Boxplot_empirical_betas}, $\bar{\beta}$
on the US stock market is persistently close to one for various time
horizons and values of $N$, consistently with the homogeneous assumption.
Taking a few technical precautions into account (\cite{brennan1975optimal}),
the differentiation of the Lagrangian (\ref{eq:functional_L}) with
respect to $N$ leads to\begin{equation}
\lambda=\frac{\overline{\mbox{Var}}(\varepsilon)P_{v}^{1-\delta}}{(1-\delta)C(1+r)\left(\frac{N^{*}}{x}\right)^{2-\delta}},\label{eq:optimal_number_of_sec_equal_weight}\end{equation}
where it is assumed that $\delta<1$ since for $\delta=1$ the optimum
investment does not depend on $N$ through the cost function. According
to (\ref{eq:optimal_number_of_sec_equal_weight}), the agent risk
tolerance increases with their account value $P_{v}$, in agreement
with various survey studies on the risk tolerance of actual investors
(see the literature review of~\cite{VanDeVenterRiskTolerance}).
Using (\ref{eq:optimal_fraction_equal_weight}) and (\ref{eq:optimal_number_of_sec_equal_weight})
to get rid of $\lambda$, we obtain\begin{equation}
N^{2-\delta}\left(1+\frac{\delta}{1-\delta}\frac{K}{N}\right)=K\frac{\bar{\beta}(E(R_{M})-r)}{(1-\delta)C(1+r)}\left(xP_{v}\right)^{1-\delta},\label{eq:N_securities_optimal_fraction}\end{equation}
where $K$ is the ratio of residual risk to market risk defined as
\begin{equation}
K=2\left(\frac{\bar{\beta}^{2}\mbox{Var}(R_{M})}{\overline{\mbox{Var}}(\varepsilon)}+\frac{1}{N}\right)^{-1}\underset{N\gg1}{\approx}2\frac{\overline{\mbox{Var}}(\varepsilon)}{\bar{\beta}^{2}\mbox{Var}(R_{M})}.\label{eq:K_ratio_definition}\end{equation}
Given the desired level of systematic risk $x$, (\ref{eq:N_securities_optimal_fraction})
can be solved for $N$ numerically in an actual portfolio optimization.
Further insight is gained by considering the high diversification
limit $N\gg1$, which yields $1+\frac{\delta}{1-\delta}\frac{K}{N}\approx1$
in (\ref{eq:N_securities_optimal_fraction}) and thus\begin{equation}
N=\left(K\frac{\bar{\beta}(E(R_{M})-r)}{(1-\delta)C(1+r)}\right)^{\frac{1}{2-\delta}}\left(xP_{v}\right)^{\frac{1-\delta}{2-\delta}},\label{eq:N_securities_optima_fraction_simplified}\end{equation}
where $K$ is given by the right-hand side of (\ref{eq:K_ratio_definition}).
The latter equation generalizes~\cite{brennan1975optimal} to the
case of a varying cost impact represented here by the parameter $\delta$
(i.e.~the result of~\cite{brennan1975optimal} is recovered by setting
$\delta=0$ and $\beta_{i}=1$ in (\ref{eq:N_securities_optima_fraction_simplified})).
These results can be further generalized to non-equally weighted portfolios
by differentiating (\ref{eq:functional_L}) with respect to $x_{i}$
and assuming again an homogeneous condition for the $\beta_{i}$s.

In essence, (\ref{eq:N_securities_optima_fraction_simplified}) says
that the number of securities held in an equally-weighted mean-variance
portfolio with Sharpe-like returns is related to the amount invested
as\begin{equation}
\log(N)=\frac{1-\delta}{2-\delta}\log(xP_{v})+\mbox{\ensuremath{\kappa}}\label{eq:log_N_log_invest}\end{equation}
in the high diversification limit, where $\kappa$ is the pre-factor
of $\left(xP_{v}\right)^{\frac{1-\delta}{2-\delta}}$ in (\ref{eq:N_securities_optima_fraction_simplified}).
The last equation gives $N$ as a function of $P_{v}$ for a predefined
$x$ in the optimal portfolio. The heterogeneity of the traders, beyond
their account value, is not apparent yet, but may occur both in $x$
and $\kappa$: first each trader may have his own preference regarding
the fraction of this account to invest in risky assets, $x$; therefore
one should replace $x$ by $x^{i}$; next, $\kappa$ includes both
a term related to transaction costs, which does vary from trader to
trader, and some measures and expectation of market returns and variance;
each trader may have his own perception or way of measuring them,
hence $\kappa$ should also be replaced by $\kappa^{i}$. Finally,
both terms can be merged in the same constant term $\zeta^{i}=\frac{1-\delta}{2-\delta}\log(x^{i})+\kappa^{i}$.
This explains how the heterogeneity of the traders is the cause of
fluctuations in the kind of relationships we are interested in.

\section{Turnover, number of assets and account value}

The result above only links $N$ with $P_{v}$, but one also wishes
to obtain relationships that involve the turnover per transaction,
$T$. Whereas in section~\ref{sec:Results}, we have characterized
the turnover of any transaction, the results of section~\ref{sec:portfolo optimization model}
rest on the assumption that the agents build their portfolio by selecting
a group of assets and stick to them over a period of time. This, obviously,
does not include the possibility of speculating by a series of buy
and sell trades on even a single asset, nor portfolio rebalancing
which consists in adjusting the relative proportions of some assets.
We thus have to find a way to differentiate between portfolio building,
rebalancing and speculation. Here, we shall focus on portfolio building
in order to test and link the results of section~\ref{sec:portfolo optimization model}
to those of section~\ref{sec:Results}.

We have found a simple effective method that can separate portfolio-building
transactions from the other ones: we assume that the transactions
of trader $i$ that correspond to the building of his portfolio are
restricted the first transaction of assets not traded previously;
sell orders are ignored, since Swissquote clients cannot short sell
easily. In other words, if trader $i$ owns some shares of assets
A, B, and C and then buys some shares of asset D, the corresponding
transaction is deemed to contribute to his portfolio building process;
the set of such transactions is denoted by $\Phi_{i}$, while the
full set of transactions is denoted by $\Omega_{i}$. Any subsequent
transaction of shares of assets A, B, C, or D are left out of $\Phi_{i}$.
The number of different assets that trader $i$ owns is supposed to
be $N_{i}\simeq|\Phi_{i}|$ where $|X|$ is the cardinal of set $X$;
this approach assumes that a trader always owns shares in all the
assets ever traded; surprisingly, this is by large the most common
case. We shall drop the index $i$ from now on.

Let us now focus on $T_{\Phi}=\sum_{k\in\Phi}T_{k}$, the total turnover
that helped building his portfolio. We should first check how it is
related to the total portfolio value $P_{v}$. Let us define $\left\langle P_{v}\right\rangle _{\Phi}$,
the account value of a trader averaged at the times at which he trades
a new asset.Plotting $\log\left\langle P_{v}\right\rangle _{\Phi}$
against $\log T_{\Phi}$ gives a cloudy relationship, as usual, but
the fitting it with $\log\left\langle P_{v}\right\rangle _{\Phi}=\chi\log T_{\Phi}$
gives $\chi=1.03\pm0.02$ for individuals, $\chi=0.99\pm0.02$ for
asset managers and $\chi=1.00\pm0.01$ for companies with an adjusted
$R^{2}=0.99$ in all cases. This relationship trivially holds for
the traders who buy all their assets at once, as assumed in the portfolio
model. The traders who do not lie on this line either hold positions
in cash (in which case this line is a lower bound), or do not build
their portfolio in a single day: they pile up positions in derivative
products or stocks whose price fluctuations are the origin of the
devations from the line. But the fact that the slope is close to 1
means that the average fluctuation is zero, hence, that on average
trades do not make money from the positions taken on new stocks. The
consequence of this is that $\log P_{v}$ can be replaced by $\log T_{\Phi}$
in (\ref{eq:log_N_log_invest}), thus, setting $x=1$,

\begin{equation}
\log N=\frac{1-\delta}{2-\delta}\log T_{\Phi}+\mbox{\ensuremath{\kappa}}\label{eq:log_N_log_Tphi}\end{equation}

The $x=1$ assumption is in fact quite reasonable: most Swissquote
traders do not use their trading account as savings accounts and are
fully invested; we do not know what amount they keep on their other
bank accounts.

\begin{figure}
\includegraphics[scale=0.65]{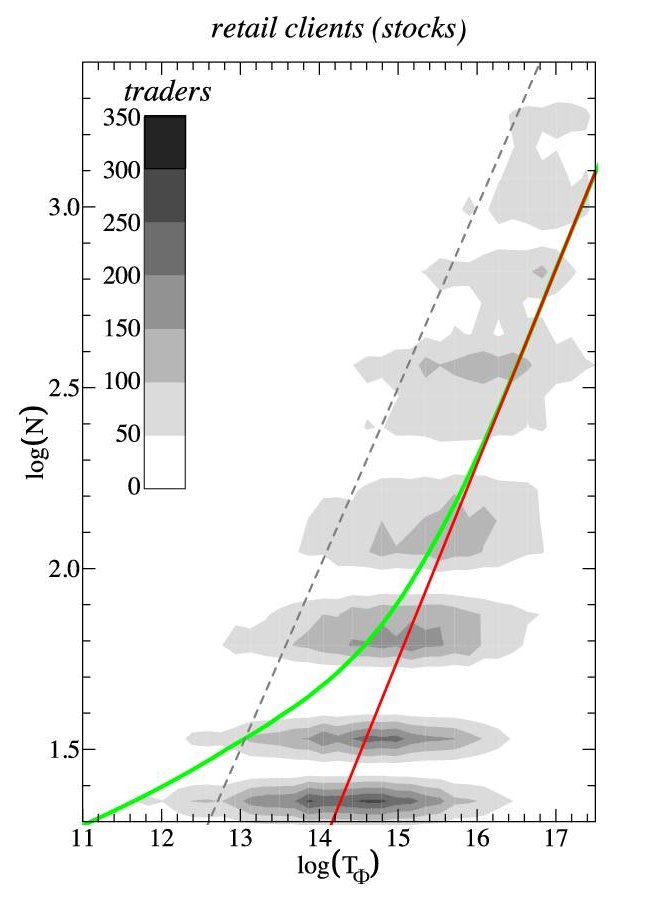}\includegraphics[scale=0.65]{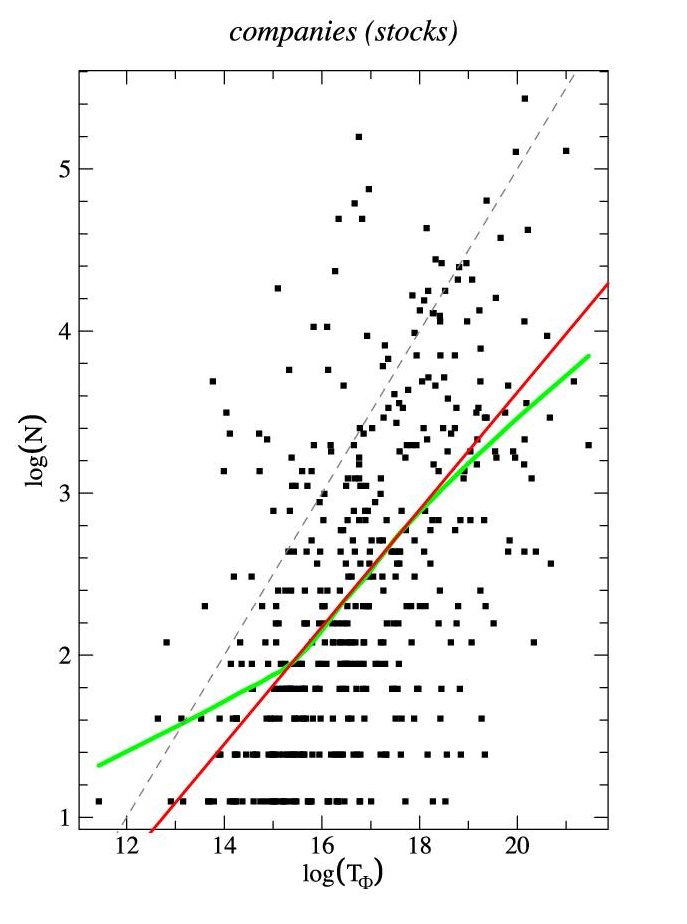}\includegraphics[scale=0.65]{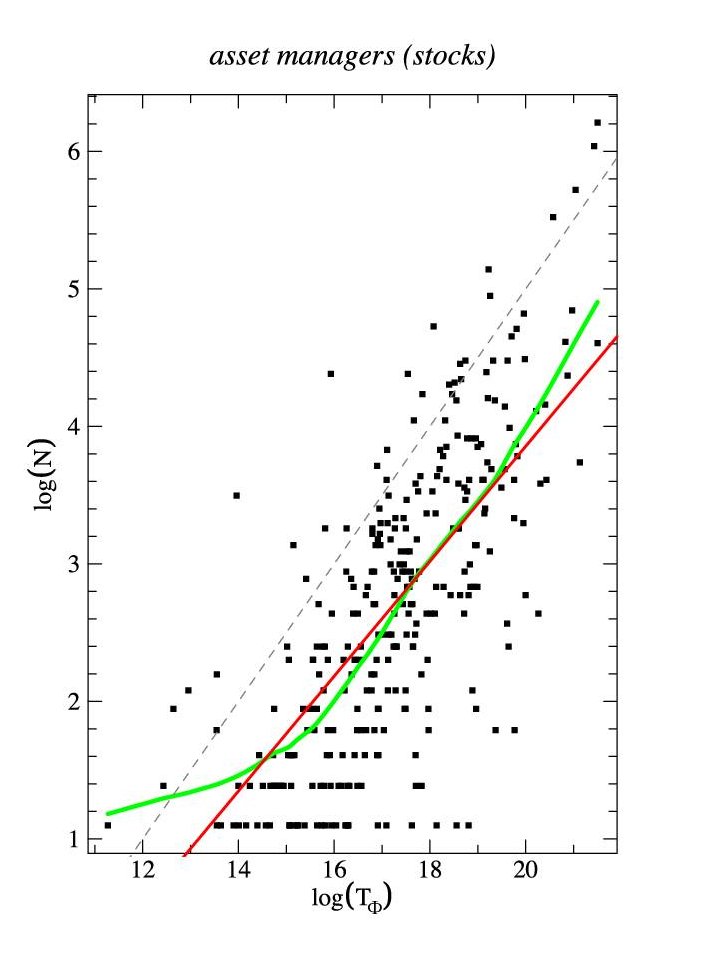}

\caption{Turnover of transactions contributing to the building of a portfolio
$T_{\Phi}$ versus the number $N$ of assets held by a given trader
at the time of the transaction. Green lines: non-parametric fit; red
lines: fits of the linear part of the non-parametric fit. From left
to right: companies, asset managers, and individuals.\label{fig:Turnover-of-transactions}}

\end{figure}

\begin{table}
\begin{centering}
\caption{Slope $\alpha$ linking $\log T_{\Phi}$ and $\log N$ for the three
trader categories.\label{tab:alpha}}

\par\end{centering}

\centering{}\begin{tabular}{cccc}
\toprule 
 & individuals & companies & asset managers\tabularnewline
\midrule
\midrule 
$\alpha$ & $0.52\pm0.02$ & $0.36\pm0.14$ & $0.44\pm0.13$\tabularnewline
\midrule 
$\log T_{\Phi}\in$ & $[16,19]$ & $[17,19.8]$ & $[15.8,18]$\tabularnewline
\bottomrule
\end{tabular}
\end{table}

\begin{table}
\caption{Results of the double linear regression of $\log\left\langle T\right\rangle _{\Phi}$
versus $\log\left\langle P_{v}\right\rangle _{\Phi}$. For each category
of investors, the first and second row correspond respectively to
$\log\left\langle P_{v}\right\rangle _{\Phi}\le\Theta_{1}$ an $\log\left\langle P_{v}\right\rangle _{\Phi}\ge\Theta_{2}$,
where $\Theta_{1,2}$ have been determined graphically using the non-parametric
method of~\cite{cleveland1988} as in section~\ref{sub:T-vs-Pv}.
Parameters are as in the double linear model (\ref{eq:T_W_relation}).
For confidentiality reasons, we have multiplied $P_{v}$ and $T$
by a random number, which only affects the true values of $\Theta_{1,2}$
and of the ordinate $a_{x}$.\label{tab:Parameters-linear-model-T-Pv-2}}

\hfill{}\begin{tabular}{lccccc}
\toprule 
 & $\beta_{x}$ & $a_{x}$ & $\xi$ & $\Theta$ & $R^{2}$\tabularnewline
\midrule
individuals & $0.85\pm0.02$ & $0.71\pm0.16$ & 0.65 & 14.5 & 0.59\tabularnewline
 & $0.51\pm0.01$ & $5.62\pm0.17$ & 0.76 & 15 & 0.31\tabularnewline
\midrule
companies & $0.83\pm0.17$ & $1.03\pm2.47$ & 0.86 & 15.5 & 0.42\tabularnewline
 & $0.62\pm0.14$ & $3.99\pm2.55$ & 0.93 & 17 & 0.32\tabularnewline
\midrule
asset managers & $0.84\pm0.25$ & $0.45\pm3.77$ & 0.79 & 15.95 & 0.50\tabularnewline
 & $0.73\pm0.17$ & $1.72\pm3.23$ & 0.72 & 18 & 0.41\tabularnewline
\bottomrule
\end{tabular}\hfill{}
\end{table}

A robust non-parametric fit does reveal a linear relationship between
$\log N$ and $\log T_{\Phi}$ in a given region $(N,T_{\Phi})\in\Gamma$
(figure~\ref{fig:Turnover-of-transactions}). In this region, we
have \begin{equation}
\log N=\alpha\log T_{\Phi}+\beta,\label{eq:2}\end{equation}
which gives \begin{equation}
\alpha=\frac{1-\delta}{2-\delta}.\label{eq:alpha_delta}\end{equation}
We still need to link $\left\langle T\right\rangle _{\Phi}$ and $\left\langle P_{v}\right\rangle _{\Phi}^{\beta}$.
While section 3 showed that the unconditional averages lead to $\left\langle T\right\rangle \sim\left\langle P_{v}\right\rangle ^{\beta}$,
one also finds that $\left\langle T\right\rangle _{\Phi}\sim\left\langle P_{v}\right\rangle _{\Phi}^{\beta}$.
Therefore, one can write\begin{equation}
\log\left\langle T\right\rangle _{\Phi}=\beta\log\left\langle P_{v}\right\rangle _{\Phi}+\mbox{cst}.\label{eq:5}\end{equation}

Thus, one is finally rewarded with the missing link \begin{equation}
\beta=\frac{1}{2-\delta},\label{eq:beta_delta}\end{equation}

which directly involves the transaction cost structure in the relationship
between turnover and portfolio value, as argued in section 3%
\footnote{Note that this relationship can be obtained directly by assuming that
all the transactions happen at the same time, hence that $T=(xP_{v})/N$,
which leads straightforwardly to (\ref{eq:beta_delta}).%
}. This relationship allows us to close the loop as we are now able
to relate directly the exponents linking $T$, $N$, and $P_{v}$.
Going back to section~\ref{sec:Results}, one understands that the
existence of a bi-linear relationship between log-turnover and log-account
value, i.e., of two values of $\beta$ for each of the three categories
of clients, is linked to two values of $\delta$: a flat flee structure
or the disregard for transaction costs leads to $\beta=\frac{1}{2}$,
while proportional fees ($\delta=1$) give $\beta=1$. 

Let us finally discuss the empirical values of $\alpha$, $\beta$,
and $\delta$ against their theoretical counterparts, which is summarized
in table \ref{tab:summary_exponents}. 
\begin{enumerate}
\item \emph{Small values of }$T_{\Phi}$: it was impossible to measure $\alpha$
in that case since the non-parametric fit shows a non-linear relationship
in the log-log plot for retail clients, which we trust more since
they have many many more points than the graphs for the two other
categories of clients. But it may not make sense to expect a linear
relationship since such a relationship is only expected for $N$ large
enough ($N\ge10$ in practice) and a small $T_{\Phi}$ is related
to a small $N$. Thus we can only test $\beta=1/(2-\delta)$. The
reported value of $\beta$ is consistent accross all the clients.
Retail clients have a larger $\delta_{eff}=2-\frac{1}{\beta}$ that
the estimated $\delta_{SQ}$. Since the shape of the fee structure
is discontinuous, the values of these exponents can hardly be expected
to match. However, fitting the whole curve structure may be problematic
in this context: indeed, the traders with a typical small value of
$T_{\Phi}$ see a more linear relationship in the region of small
transaction value that when considering the whole curve; for instance,
removing the two largest segments from the fee structure yields $\delta_{SQ}'=0.74\in[0.43,0.79]$,
which is not far of $\delta_{eff}$. 
\item \emph{Large values of} $T_{\Phi}$: the relationships between all
the exponents are verified for the three categories of clients. While
not very impressive for companies and asset managers, this result
is much stronger in the case of retail clients since the relative
uncertainties associated with each measured exponent are small (1-2\%).
The value of $\beta_{retail}$ is of particular interest as it corresponds
$\delta_{eff}=0$, or equivalently, to a flat fee structure. Going
back to the fees structure of Swissquote, one finds that that the
transition happens when the relative transaction cost falls below
some threshold (we cannot give its precise value for confidentiality
reasons; it is smaller than 1\%). A possible explanation is that either
some traders with a high enough average turnover have a flat-fee agreement
with Swissquote and that the rest of them simply act as if they were
not able to take correctly into account transaction costs. Since not
all traders have a flat-fee aggrement, one must conclude that some
traders have indeed some problems estimating small relative fees and
simply disregard them. The reported value of $\beta$ for companies
and asset managers is larger that $\beta_{retail}$, but it is more
likely than not that the small sample size is responsible for this
discrepancy, since these two categories of clients have a greater
propensity to negociate a flat-fee structure. 
\item \emph{Transition between the two regimes:} the transitions between
the standard Swissquote and an effective flat-fee structure happens
occur \emph{at the same average value} of $T$ for the three categories
of traders (idem for $T_{\Phi}$). Since there is no automatic switching
between fee structures at Swissquote for any predefined value of transaction
value, one is lead to conclude that this transition has behavioural
origins, which is also responsible for the value at which the transition
takes place which, in passing, corresponds to the end of the plateau
of the RCDF of $P_{v}$ in the case of retail clients ($e^{15}\simeq3.27\cdot10^{6}$).
As a consequence, it is likely that the traders tend to either neglect
or consider as constant transaction fees smaller than some threshold
when they build their portfolio.
\end{enumerate}
\begin{table}
\caption{\label{tab:summary_exponents}Table summarising the empirical and
theoretical relationships between $\alpha$, $\beta$, and $\delta$.}

\begin{centering}
\begin{tabular}{cccc}
\toprule 
small $T_{\phi}$ & individuals & companies & asset managers\tabularnewline
\midrule
$\beta$ & $0.85\pm0.02$ & $0.83\pm0.17$ & $0.84\pm0.25$\tabularnewline
$\log T_{\Phi}<$ & 14.5 & 17 & 18\tabularnewline
\midrule
$\delta_{eff}=2-\frac{1}{\beta}$ & $0.82\pm0.02$ & $0.80\pm0.20$ & $0.81\pm0.30$\tabularnewline
\midrule 
$\delta_{SQ}$ & $0.63\in[0.50,0.74]$ & $0.63\in[0.50,0.74]$ & $0.63\in[0.50,0.74]$\tabularnewline
$\delta_{SQ}'$ & $0.74\in[0.43,0.79]$ & $0.74\in[0.43,0.79]$ & $0.74\in[0.43,0.79]$\tabularnewline
\midrule
$\tilde{\beta}=\frac{1}{2-\delta_{SQ}}$ & $0.73\in[0.66,0.74]$ & $0.73\in[0.66,0.74]$ & $0.73\in[0.66,0.74]$\tabularnewline
\bottomrule
\end{tabular}
\par\end{centering}

\bigskip{}

\centering{}\begin{tabular}{cccc}
\toprule 
large $T_{\phi}$ & individuals & companies & asset managers\tabularnewline
\midrule
$\beta$ & $0.51\pm0.01$ & $0.62\pm0.14$ & $0.73\pm0.17$\tabularnewline
$\log T_{\Phi}>$ & 15 & 17 & 18\tabularnewline
\midrule
$\delta_{eff}=2-\frac{1}{\beta}$ & $0.04\pm0.02$ & $0.39\pm0.23$ & $0.63\pm0.23$\tabularnewline
$\alpha_{eff}=\frac{1-\delta_{eff}}{2-\delta_{eff}}$ & $0.49\pm0.01$ & $0.38\pm0.09$ & $0.27\pm0.08$\tabularnewline
\midrule
$\alpha$ & $0.52\pm0.02$ & $0.36\pm0.14$ & $0.44\pm0.13$\tabularnewline
$\log T_{\Phi}\in$ & $[16,19]$ & $[17,19.8]$ & $[15.8,18]$\tabularnewline
\bottomrule
\end{tabular}
\end{table}

{}

\section{Discussion and outlook}

We have been able to determine empirically a bilinear relationship
between the average log-turnover and the average log-account value
and have argued that it comes from the transaction fee structure of
the broker and its perception by the agents. A theoretical derivation
of optimal simple one-shot mean-variance portfolios with non-linear
transaction costs predicted relationships between turnover, number
of different asset in the portfolio and log-account values that could
be verified empirically. This means that the populations of traders
do take correctly \emph{on average, i.e. collectively,} the transaction
costs into account and act $collectively$ as mean-variance equally-weighted
portfolio optimizers. This is not to say that each trader is a mean-variance
optimizer, but that the population taken as a whole behaves as such---with
differences across populations, as discussed in the previous section.
This to be related to findings of Kirman's famous work on demand and
offer average curves in Marseille's fish market~\cite{hardle1995nonclassical}
and more generally as what has become known as the wisdom of the crowds
(see~\cite{wisdomcrowds} for an easy-to-read account). 

The fact that the turnover depends in a non-linear way on the account
value implies that linking the exponents of the distributions of transaction
volume, buying power of large players in financial markets, and price
return is more complex that previously thought \cite{Gabaix_etalNature2003}.
It has also implications for agent-based models, which from now on
must take into account the fact that the real traders do invest into
a number of assets that depends non-linearly on their wealth.

Future research will address the relationship between account value
and trading frequency, which is of utmost importance to understand
if the many small trades of small investors have a comparable influence
on financial market than those of institutional investors. This will
give an understanding of whom provides liquidity and what all the
non-linear relationships found above mean in this respect. This is
also crucial in agent-based models, in which one often imposes such
relationship by hand, arbitrarily; reversely, one will be able to
validate evolutionary mechanisms of agent-based model according to
the relationship between trading frequency, turnover, number of assets
and account value they achieve in their steady state.

\bibliographystyle{acm}
\bibliography{biblio}

\end{document}